\documentclass[fleqn,usenatbib]{mnras}
\usepackage{xcolor}     
\usepackage{tikz}       
\usepackage{pgffor}     
\usepackage{hyperref}   
\usepackage{placeins}

\definecolor{lime}{HTML}{A6CE39}
\DeclareRobustCommand{\orcidicon}{%
	\begin{tikzpicture}
	\draw[lime, fill=lime] (0,0) 
	circle [radius=0.16] 
	node[white] {{\fontfamily{qag}\selectfont \tiny ID}};
	\draw[white, fill=white] (-0.0625,0.095) 
	circle [radius=0.007];
	\end{tikzpicture}
	\hspace{-2mm}
}

\foreach \x in {A, ..., Z}{%
	\expandafter\xdef\csname orcid\x\endcsname{\noexpand\href{https://orcid.org/\csname orcidauthor\x\endcsname}{\noexpand\orcidicon}}
}

\usepackage{newtxtext,newtxmath}

\usepackage[T1]{fontenc}

\DeclareRobustCommand{\VAN}[3]{#2}
\let\VANthebibliography\thebibliography
\def\thebibliography{\DeclareRobustCommand{\VAN}[3]{##3}\VANthebibliography}

\usepackage{graphicx}	
\usepackage{amsmath}	

\title[GW Probes of Globular Clusters]{Probing Globular Clusters using Gravitational Waves from Inspiraling Stellar-mass Binary Black Holes in Space-based Detector}

\author[J. Wu et al.]{
Jie Wu\orcidA{}$^{1,2}$,
Yao Xiao\orcidB{}$^{3}$,
Mengfei Sun\orcidC{}$^{1,2}$
and Jin Li\orcidD{}$^{1,2,4}$\thanks{E-mail: cqujinli1983@cqu.edu.cn}
\\
$^{1}$College of Physics, Chongqing University, Chongqing 401331, China\\
$^{2}$Department of Physics and Chongqing Key Laboratory for Strongly Coupled Physics, Chongqing University, Chongqing 401331, China\\
$^{3}$South-Western Institute for Astronomy Research, Yunnan University, Kunming 650500, China\\
$^{4}$Institute of Advanced Interdisciplinary Studies, Chongqing University, Chongqing 401331, China}

\date{Accepted 2026 September 28. Received 2026 September 28; in original form 2025 December 23}

\pubyear{2026}
\volume{552}
\journal{\parbox{\textwidth}{%
MNRAS \textbf{552}, \pageref{firstpage}--\pageref{lastpage} (2026)\hfill
\href{https://doi.org/10.1093/mnras/stag1891}{https://doi.org/10.1093/mnras/stag1891}\\
Advance Access publication 2026 October 7}}

\begin{document}
\label{firstpage}
\pagerange{\pageref{firstpage}--\pageref{lastpage}}
\maketitle

\begin{abstract}
Globular clusters (GCs) serve as key environments for studying stellar dynamics and galactic structure, yet precise measurements of their distances and masses are often limited by uncertainties in electromagnetic (EM) observations.
We present a method that uses gravitational waves (GWs) from inspiralling stellar-mass binary black holes (BBHs) orbiting within GCs to improve the precision of GC parameter measurements.
The BBH’s circular motion around the GC produces a moving-source effect, including Doppler shifts and relativistic corrections, which leaves measurable imprints on the GW signal.
We model these signals using post-Newtonian waveforms and apply Lorentz transformations to compute the waveform from the moving BBHs.
The information encoded in the modulated waveform is quantified with the Fisher information matrix, allowing us to extract GC parameters and combine them with EM measurements.
Our results show that incorporating GW observations can substantially improve distance constraints for nearby Milky Way GCs, while the improvement in mass constraints depends strongly on the BBH outer-orbit modulation.
These findings demonstrate the value of BBHs as dynamical probes and highlight the role of GWs as complementary probes of GC properties, especially for nearby and strongly modulated systems.
\end{abstract}




\section{Introduction}
Globular clusters (GCs) are among the oldest stellar systems in the Universe. 
They typically consist of dense, gravitationally bound collections of about $10^4-10^6$ stars, serving as important laboratories for studying stellar evolution, galaxy formation, and cosmology~\citep{Benacquista2013}.
Because of their longevity and widespread presence in galaxies, GCs provide critical insights into the formation histories of their hosts, making them key tracers of galaxy evolution and cosmic structure assembly~\citep{Forbes2018,Meylan1997}.

Precise measurements of GC distances and masses are of particular importance.
Accurate determinations of GC distances are fundamental for establishing the astronomical distance scale.
They serve as valuable calibration points for standard candles and help improve extinction corrections within the Milky Way and nearby galaxies~\citep{Rejkuba2012,Chaboyer1999}.
High-precision GC distance measurements thus play a vital role in mapping cosmic structures, refining the cosmic distance ladder, and constraining cosmological parameters such as the Hubble constant~\citep{MWGC,Riess2021}.
Likewise, accurately determining GC masses is essential for understanding their internal dynamics, formation mechanisms, and subsequent evolution.
Mass measurements reveal core collapse phenomena, mass segregation, and stellar interactions within GCs, thereby refining theoretical models of their long-term evolution~\citep{Baumgardt2017}.
Furthermore, GC mass distributions provide insight into dark matter content and stellar populations, highlighting the interplay between internal dynamics and the external galactic environment~\citep{Hudson2014,Harris2017}.

Improving the precision of GC distance and mass measurements remains an active topic in astrophysics.
However, current methods for determining GC distances and masses primarily rely on electromagnetic (EM) observations.
Distances are commonly determined using standard candles (e.g., RR Lyrae stars and Type II Cepheids) or via trigonometric parallaxes~\citep{Bhardwaj2022,Vasiliev2021}.
Masses are typically inferred from dynamical models based on velocity dispersion measurements or surface brightness profiles~\citep{Zocchi2012,Baumgardt2009}.
Although these techniques have led to significant progress, they remain limited by factors such as dust extinction, line-of-sight projection effects, and assumptions of dynamical equilibrium~\citep{Majaess2012}.
Despite extensive efforts, these systematics continue to limit distance and mass estimates for many GCs.
These limitations motivate the development of alternative and independent approaches to improve the precision of GC parameter measurements.

Since the first detection of gravitational waves (GWs) in 2015, GW astronomy has rapidly grown into a transformative field, offering a fundamentally different observational perspective compared to traditional EM methods~\citep{GW150914,GW_Phys}.
A major advantage of GW signals is their immunity to dust extinction and line-of-sight projection effects, enabling robust detection even in crowded or obscured regions~\citep{Hirata2010,Colaco2024}.
Among various GW sources, stellar-mass binary black holes (BBHs) stand out as the most prominent and best-studied population to date~\citep{Gerosa2021}.
Most GW events detected so far originate from BBH mergers, making BBHs the most common and observationally accessible GW sources~\citep{GWTC3}.
The waveform properties of BBHs have been extensively analyzed for a broad range of masses and orbital configurations, providing a solid foundation for extracting astrophysical parameters from GW signals~\citep{Schmidt2020}.
These properties make BBHs powerful tools for providing independent constraints on GC distances and masses via GW observations.
At the population level, GW observations have also been proposed as probes of GC formation and evolution~\citep{RomeroShaw2021}.

The potential of using BBHs to probe GCs has already been explored in several earlier studies, laying the groundwork for subsequent developments.
\citet{Rodriguez2015} and \citet{Kremer2018} investigated the detectability of BBHs originating in GCs using ground-based facilities (e.g., LIGO) and space-based observatories (e.g., LISA).
Their results confirmed that such systems are likely to be detectable with both current and future detectors.
\citet{Strokov2022} proposed measuring the Doppler shift induced by a BBH's radial motion as a method to detect intermediate-mass black holes in clusters.
\citet{Tiwari2024} showed that the line-of-sight acceleration of a BBH (detectable by space-based detectors such as DECIGO and LISA) can be used to infer the physical properties of its host cluster.
\citet{TorresOrjuela2025} further showed that motion induced by the surrounding stellar environment in GCs can leave measurable imprints on the GW phase.
In a related study, \citet{Stegmann2024} found that when stellar-mass BBHs orbit supermassive black hole binaries (SMBHBs), the BBHs' motion can imprint characteristic modulations on decihertz GW signals. 
These modulation features enable population studies of SMBHBs through detailed analyses of the waveform structure.
\citet{Tiwari2025Env} argued more broadly that GW signals can be used to profile the stellar environments of compact binaries.
In a related application, \citet{Wu_AGN} considered using inspiraling stellar-mass BBHs to constrain the properties of accretion disks surrounding active galactic nuclei.
The same physical principle applies to BBHs embedded in GCs.
Collectively, these studies exploit how the orbital motion of BBHs around massive objects influences GW signals, underscoring a promising approach to probing cluster parameters via detailed waveform analysis.

In this work, we present a new GW-based method for probing GCs.  
Specifically, we examine how the orbital motion of a stellar-mass BBH within a GC modulates its GW signal through waveform transformations derived from Lorentz boosts. 
These transformations capture the effects of BBHs orbiting within the GC and enable constraints on the host GC's distance and mass.
By applying a Fisher information matrix (FIM) analysis to GW signals, we quantify the uncertainties in GC parameters achievable with LISA observations and compare these with existing EM constraints. 
This allows us to assess the improvement from combining GW and EM measurements in a multi-messenger framework.
A key innovation of our approach is the implementation of full waveform transformations that account for arbitrary source motion directions, going beyond previous treatments that were limited to line-of-sight velocity and edge-on orbital configurations, neglecting the effect of transverse motion.    
The GW signals used in this study are simulated using second-generation time-delay interferometry (TDI), ensuring consistency with future space-based GW observations.
Furthermore, we construct asymmetric normal distributions based on existing EM observations to model the combined parameter distributions from EM and GW data, allowing us to assess the theoretical potential of multi-messenger constraints.  
Overall, these developments offer a new perspective and methodology for GC studies.

The structure of this paper is organized as follows.
In Sec.~\ref{sec:Methods}, we describe the adopted GW waveform model, the moving-source transformation, the GC data, and the overall data analysis procedure.  
In Sec.~\ref{sec:Results}, we present how incorporating modulated GWs improves the constraints on GC parameters relative to EM observations.  
Finally, Sec.~\ref{sec:Discussion} discusses several additional details.  
Appendix~\ref{app:emdata} summarizes the EM GC data, while Appendix~\ref{app:plummerking} compares the Plummer and King profiles used in this work.
Throughout this paper, we use geometric units with $c = G = 1$.

\section{Methods}\label{sec:Methods}
\subsection{GW signal}

In general relativity, GWs are tensorial perturbations with boost weight zero and spin weight two in four-dimensional spacetime, which makes fully covariant Lorentz transformations nontrivial~\citep{Cao_GWtrans,Cao_GWtrans_2026}. 
In the transverse–traceless description, the GW can be represented by a three-dimensional spatial tensor,
\begin{equation}
  h_{ij}=h_{+}e^{+}_{ij}+h_{\times}e^{\times}_{ij},
\end{equation}
where $h_{+}$ and $h_{\times}$ are the plus and cross polarizations and $e^{+}_{ij}$, $e^{\times}_{ij}$ are the corresponding polarization tensors~\citep{Wu_TDI}. 
This representation permits frame changes to be carried out with three-dimensional Lorentz transformations supplemented by a rescaling of the time coordinate.
A complete derivation of this formalism can be found in \citet{Cao_GWtrans,Cao_GWtrans_2026}.
Under this approach, the GW signal $h_{ij}$ in a frame moving at an arbitrary velocity $\vec{v}\text{(t)}$ relative to the rest frame can be expressed as
\begin{equation}
	\begin{aligned}
		h_{ij}' & =h_{ij}+v^{k}h_{kl}v^{l}\frac{1}{(1-\hat{r}\cdot\vec{v})^{2}}[\hat{r}_{i}\hat{r}_{j} \\
		 & -\frac{\gamma}{1+\gamma}(\hat{r}_{i}v_{j}+v_{i}\hat{r}_{j})+\frac{\gamma^{2}}{(1+\gamma)^{2}}v_{i}v_{j}] \\
		 & +v^{k}h_{kj}\frac{1}{1-\hat{r}\cdot\vec{v}}[\hat{r}_{i}-\frac{\gamma}{1+\gamma}v_{i}] \\
		 & +v^{k}h_{ik}\frac{1}{1-\hat{r}\cdot\vec{v}}[\hat{r}_{j}-\frac{\gamma}{1+\gamma}v_{j}], 
		\end{aligned}
\end{equation}
with
\begin{equation}
\gamma(t) = \frac{1}{\sqrt{1-v(t)^{2}}}.
\end{equation}
The corresponding transformation of the time coordinate can be expressed as  
\begin{equation}
\begin{aligned}
t'(t)=\int_0^t
\frac{d\tau}{\gamma(\tau)\left[1-\vec{v}(\tau)\cdot\hat{r}(\tau)\right]},
\end{aligned}
\label{eq:time_transform}
\end{equation}
where $t$ is the observer's time and $t'$ is the time of the isolated-BBH waveform.
Although the Lorentz transformation above is defined between inertial frames, an accelerating source can be treated as a sequence of instantaneous comoving inertial frames, with the time-dependent velocity $\vec{v}(t)$ entering the waveform through the local boost and the integral time mapping~\citep{Cao_GWtrans_2026}.
This adiabatic local-boost prescription is appropriate here because the BBH centre-of-mass motion around the GC changes on the outer-orbit time-scale, whereas the GW carrier is generated by the much shorter inner-binary dynamics.
This formulation enables us to incorporate the velocity arising from the BBH's orbital motion around the GC into the waveform.
In the numerical implementation, we first construct the waveform in the time domain and apply the time-dependent transformation through Eq.~\eqref{eq:time_transform} before Fourier transforming the signal. The corresponding Doppler frequency shift is therefore included in the waveform.

In this work, we focus on GWs from stellar-mass BBHs within the LISA sensitivity band ($\sim 1-100\,\mathrm{mHz}$). 
For such systems, the GW emission corresponds to the early inspiral phase. 
These binaries are typically tens to thousands of years away from merger, and their inner orbits evolve slowly over observational time-scales. 
As a result, the post-Newtonian (PN) method is particularly suitable for accurate modelling.
We adopt the 3.5PN formalism, including both spin–orbit and spin–spin interactions, to construct the inspiral waveforms in the BBH rest frame~\citep{Wu_PN,PN1,PN2}. 
This level of PN accuracy ensures consistent phase evolution across the multi-year LISA observation window and provides sufficient fidelity for evaluating modulation effects induced by the BBH’s outer orbit within the host GC.

\begin{figure}
    \begin{minipage}{\columnwidth}
        \centering
        \includegraphics[width=0.9\textwidth,
        trim=0 0 0 0,clip]{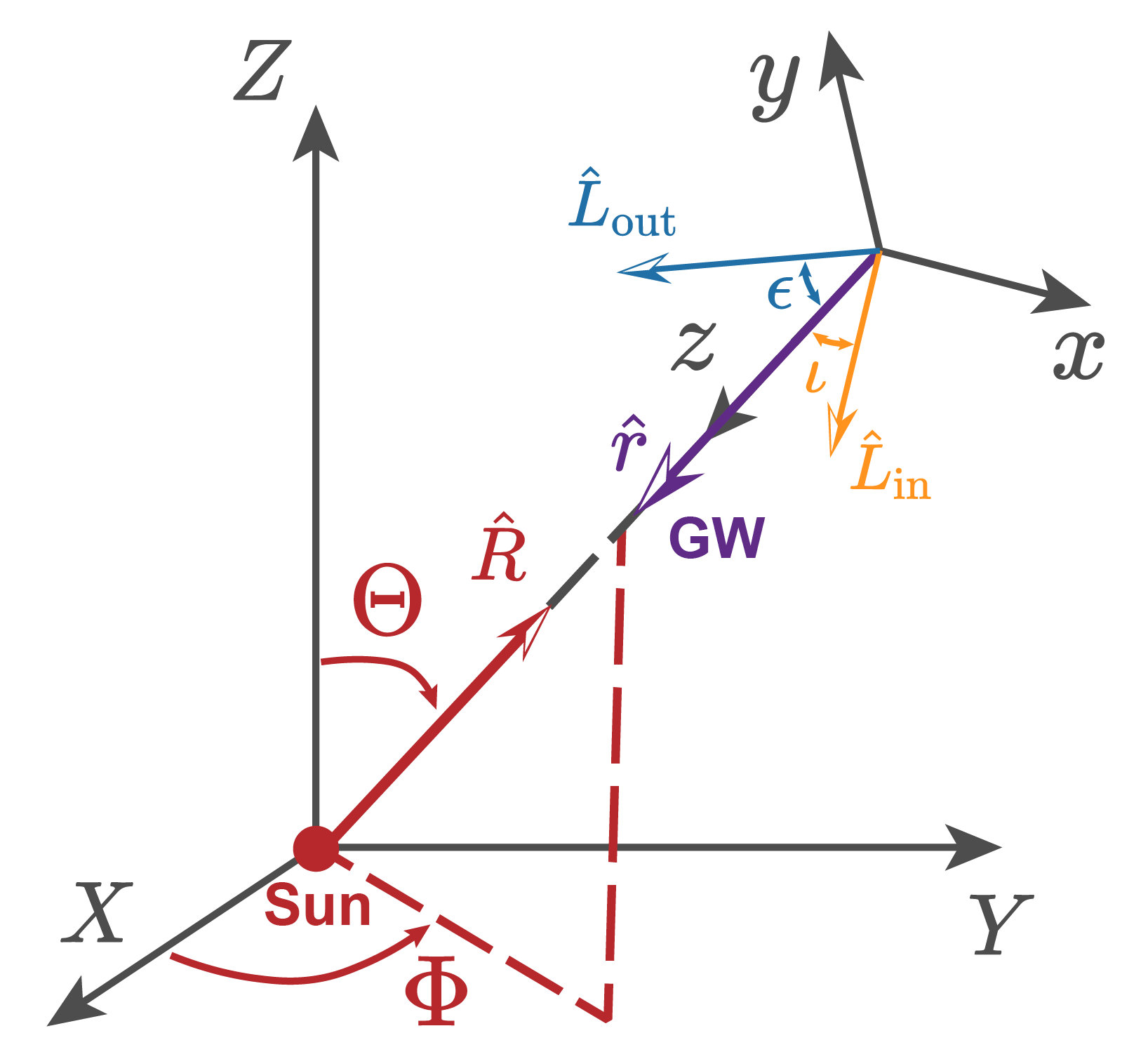}
        \caption{Schematic illustration of the SSB frame $(X,Y,Z)$ and the GC frame $(x,y,z)$. The BBH forms an inner binary, while its centre of mass follows an outer orbit around the GC (see Fig.~\ref{fig:rest_moving}(b)). The angles between the corresponding orbital angular momentum vectors $\hat{L}$ and the GW propagation direction $\hat{r}$ are denoted by $\iota$ (inner orbit) and $\epsilon$ (outer orbit). The transformation between the SSB frame and the equatorial coordinate system adopts the obliquity of the ecliptic ($\varepsilon_{2000} = 23^{\circ}\,26'\,21''$) based on epoch J2000.0~\citep{obliquity_of_ecliptic}.}\label{fig:coordinates}
    \end{minipage}
\end{figure}

Following the PN waveform construction and the moving-source transformation, we compute the LISA detector response to the modulated signal in the solar-system barycentric (SSB) frame.  
The BBH's orbital velocity around the GC is first evaluated in its local orbital frame and then rotated into the SSB frame using a sequence of Euler rotations.  
The geometric configuration of the BBH, its orbital angular momentum, and the relative orientation between the GC frame and the SSB frame are illustrated in Figs.~\ref{fig:coordinates} and \ref{fig:rest_moving}.  

Once the source-frame quantities are projected into the SSB frame, the GW strain is evaluated along the time-dependent arm geometry of the LISA constellation. 
We adopt second-order eccentric Keplerian orbits for the three spacecraft, allowing for time-dependent and unequal arm lengths that naturally arise as the constellation moves around the Sun~\citep{LISA_orbit}. 
From the SSB-frame metric perturbation, we project the strain onto the time-varying arm geometry to obtain the raw inter-spacecraft measurements.

Future space-based GW missions are expected to be dominated by laser-frequency noise at the raw measurement level. 
To suppress this noise, we employ second-generation TDI combinations, which form noise-cancelled Michelson-type data streams in the $X$, $Y$, and $Z$ channels.  
These channels constitute the final data products used for Fourier-domain analysis and for evaluating the FIM in our parameter-estimation pipeline.  
The source-dependent TDI response retains the sky location and polarization angles of each BBH in the waveform and FIM.
For the detector-noise model, we use the analytic one-sided power spectral density (PSD) following \citet{Babak2021LISASensitivity}.
A full description of the response model, arm-projection operators, and TDI implementation can be found in \citet{Wu_TDI}.

The procedures outlined above complete the construction of the GW signal as measured by LISA in the presence of source motion.  
Starting from the PN waveform in the BBH rest frame, we incorporate the effects of the outer orbital motion through a moving-source transformation, rotate all source-frame quantities into the SSB frame, and propagate the resulting strain through the time-dependent detector geometry.  
The final TDI observables therefore encode both the intrinsic inspiral waveform and the GC-induced modulation, providing the basis for the parameter-estimation analysis presented in the following sections.

\subsection{GW modulation}

\begin{figure}
    \begin{minipage}{\columnwidth}
        \centering
        \includegraphics[width=0.97\textwidth,
        trim=0 0 0 0,clip]{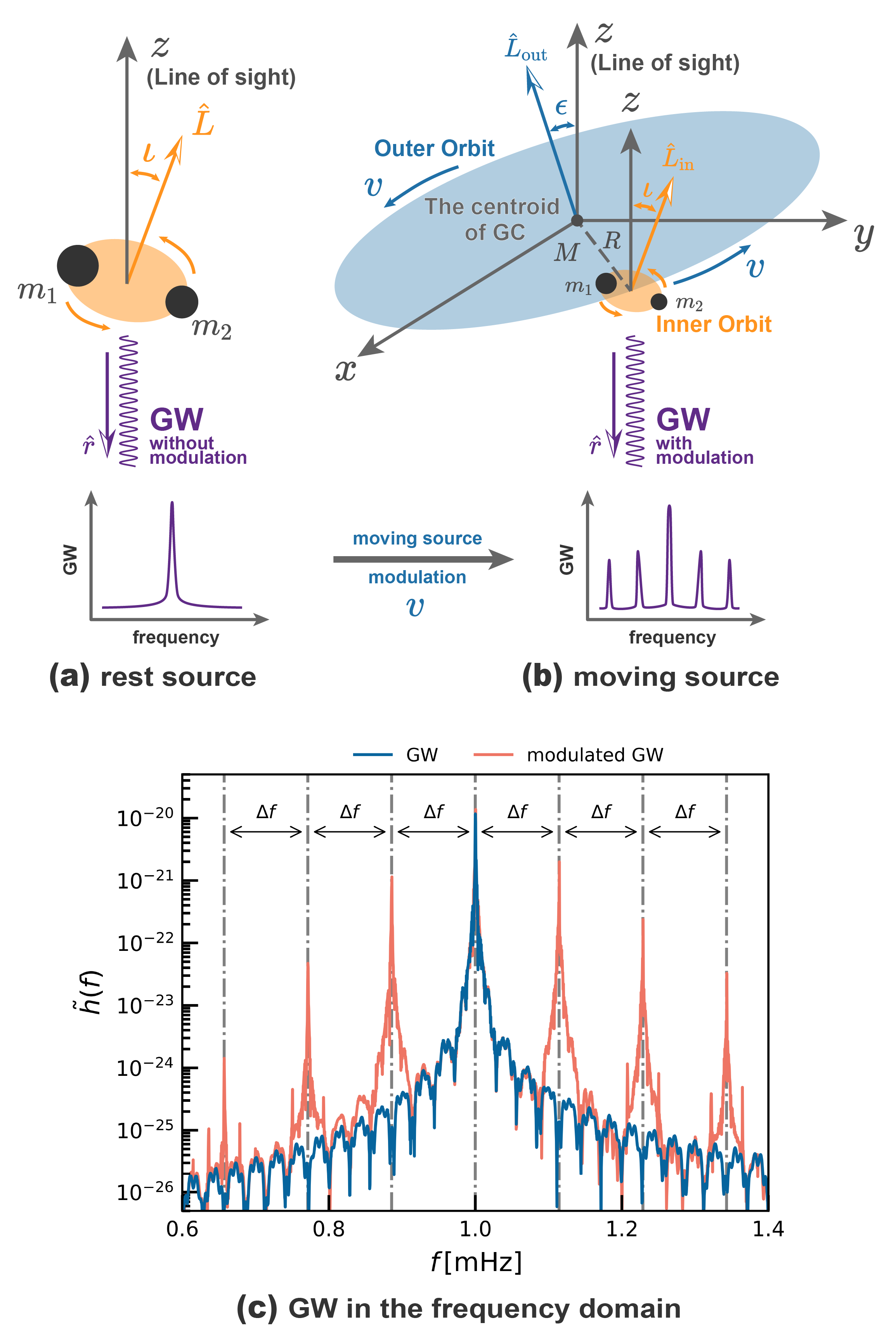}
        \caption{Schematic illustration of GW modulation. Panel (a) depicts a BBH at rest with respect to the observer, emitting an unmodulated GW signal. Panel (b) shows a BBH moving in an outer orbit around the GC's centroid, which introduces periodic modulation in the observed GW signal. In both cases, the lower sub-panels present simplified representations of the corresponding GW frequency spectrum. Panel (c) compares the frequency-domain waveform before and after modulation for a representative GW frequency of $1\,\mathrm{mHz}$, where the induced sideband spacing is $\Delta f \approx 0.1\,\mathrm{mHz}$.}\label{fig:rest_moving}
    \end{minipage}
\end{figure}

For BBH systems in the rest frame, the GWs observed by a space-based detector like LISA correspond to the early inspiral phase, where the intrinsic waveform evolves slowly over multi-year time-scales. In the calculation below, however, we retain the full 3.5PN phase evolution rather than treating the signal as strictly monochromatic. 
When the BBH is embedded in a GC and follows an outer orbit around its centre of mass, the velocity produces time-dependent Doppler shifts and relativistic corrections in the GW waveform. 
These moving-source effects introduce periodic modulations that are absent in the rest-frame waveform and leave distinct signatures in the frequency domain.

As illustrated in Fig.~\ref{fig:rest_moving}, the waveform of a moving BBH source differs significantly from that of a rest BBH. 
In the frequency domain, the outer-orbit motion generates a family of sidebands around the intrinsic carrier frequency. 
These sidebands are separated by integer multiples of the outer orbital frequency, producing a regular spacing that directly reflects the dynamical time-scale of the orbit. 
The spacing is determined by the outer orbital period $P_{\mathrm{out}}$,  
\begin{equation}
\Delta f = \frac{1}{P_{\mathrm{out}}} 
= \frac{1}{2\pi}\sqrt{\frac{M_{\rm enc}}{R^{3}}},
\end{equation}
where $M_{\rm enc}$ is the mass enclosed within the BBH outer orbit and $R$ is the outer orbital radius. The relation between $M_{\rm enc}$ and the total GC mass is specified in Sec.~\ref{subsec:priori}.
For the retained systems considered in this work, the resulting modulation frequency remains small and satisfies $\Delta f \ll f_0$. 
The example shown in Fig.~\ref{fig:rest_moving}(c) uses exaggerated parameters ($f_0 = 1\,\mathrm{mHz}$, $\Delta f = 0.1f_0$) solely to make the sideband structure visually identifiable and does not represent the parameter range used in our analysis.
For most systems considered here, the BBH outer orbit period is longer than the observation time, so the velocity evolves only slowly during the observation.
Nevertheless, GW phase measurements are highly sensitive to small frequency changes.
The time-dependent velocity components can therefore accumulate Doppler and relativistic corrections in the observed signal, allowing these weak effects to be resolved.
The associated parameter degeneracies are treated through the enclosed-mass prescription and ratio prior introduced in Sec.~\ref{subsec:priori}.

The modulation therefore provides an additional channel for extracting information from the GW signal.
Unlike the intrinsic BBH parameters, which primarily affect the carrier waveform, the GC-induced modulation encodes the BBH centre-of-mass motion in the host potential. 
Analysis of these modulation features enables us to infer key GC parameters, in particular the GC's distance and mass.
Moreover, because the modulation scales with the outer orbit's velocity and geometry, the method is naturally sensitive to the dynamical environment rather than to BBH properties alone.
This motivates the parameter choices adopted in the following section, where we specify the quantities that enter our waveform model and are subsequently constrained through the FIM analysis.

\FloatBarrier
\subsection{Parameter selection}\label{subsec:param_selection}
\begin{figure}
    \begin{minipage}{\columnwidth}
        \centering
        \includegraphics[width=0.96\textwidth,
        trim=0 0 0 0,clip]{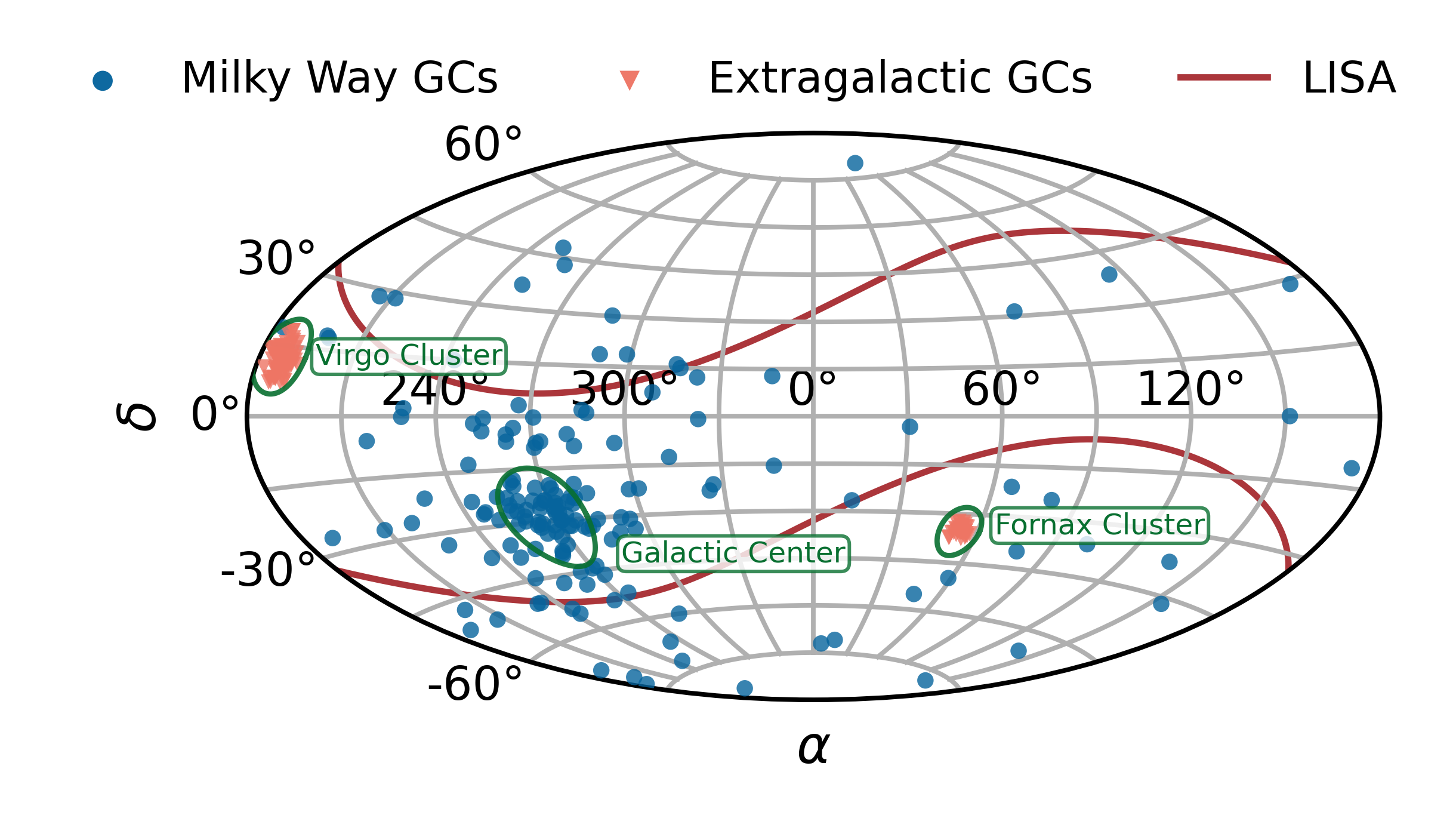}
        \caption{Sky distribution of the EM-observed GCs used in this work. Circles indicate Milky Way GCs, and inverted triangles indicate extragalactic GCs. The solid curve shows the annual trajectory of the normal vector of the LISA constellation plane.}\label{fig:location}
    \end{minipage}
\end{figure}
The GC parameters adopted in this study are drawn from two EM samples: a Milky Way GC sample based on the Galactic GC catalogue of~\citet{MWGC}, and an extragalactic GC sample constructed from ACS Virgo Cluster Survey (ACSVCS) and ACS Fornax Cluster Survey (ACSFCS) photometry with host galaxy surface brightness fluctuation (SBF) distances~\citep{Jordan2009ACSVCS,Jordan2015ACSFCS,Blakeslee2009SBF}.
From these data, we extract key GC properties, including the GC mass $M$, luminosity distance $D_L$, and sky location (right ascension $\alpha$ and declination $\delta$).
Further details of the EM data are given in Appendix~\ref{app:emdata}.
The spatial distribution of the selected GCs is illustrated in Fig.~\ref{fig:location}, with Milky Way GCs clustering around the Galactic Center and extragalactic GCs located primarily in the Virgo and Fornax clusters.
This wide coverage ensures a diverse sample across different GC environments and distance scales.

\begin{figure}
    \begin{minipage}{\columnwidth}
        \centering
        \includegraphics[width=0.96\textwidth,
        trim=0 0 0 0,clip]{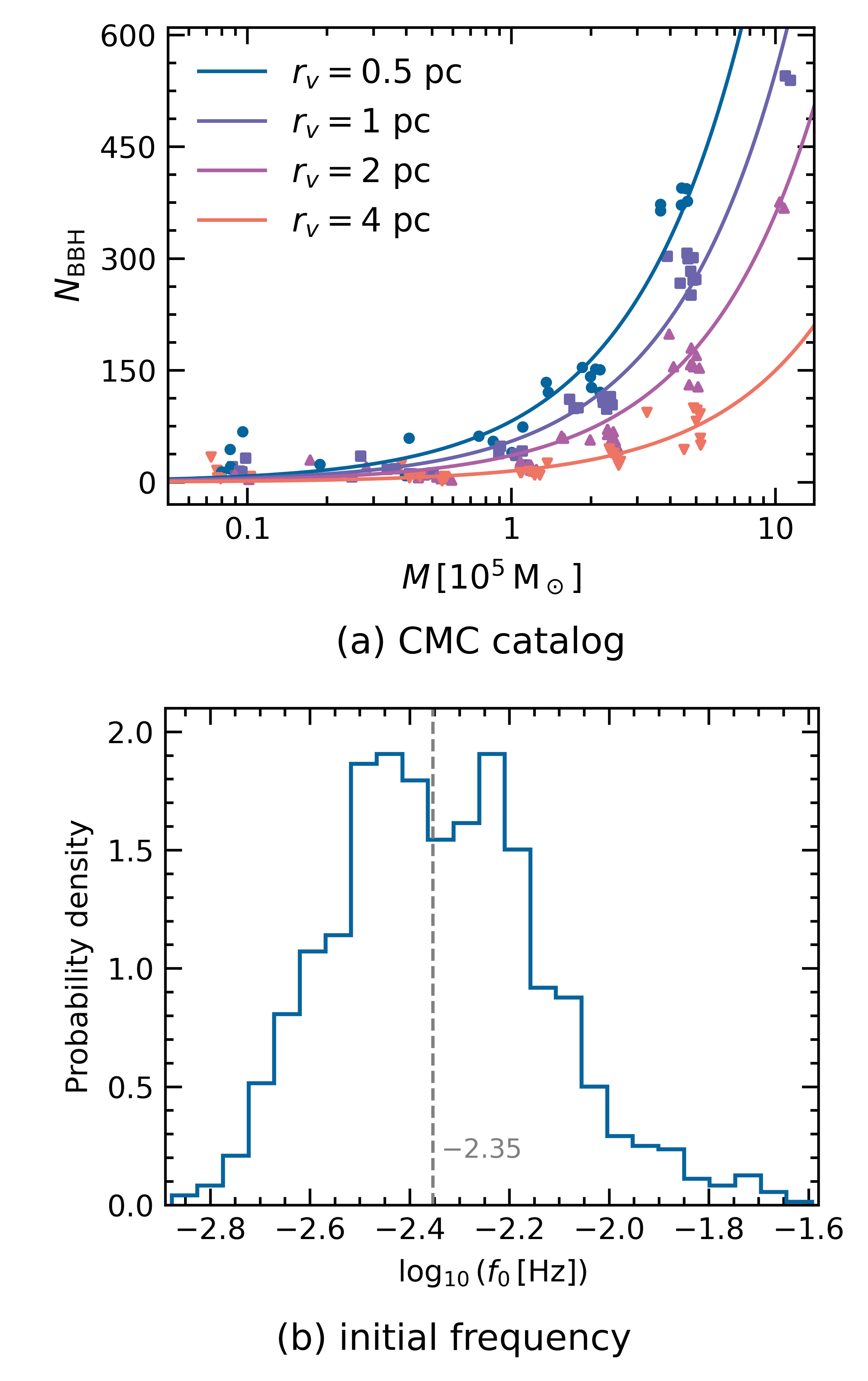}
        \caption{Properties of the CMC BBH population used in this study. Panel (a) presents the number of BBHs contained in CMC clusters of different masses, separated by the virial radius $r_v$. Solid lines denote linear fits to the data for each $r_v$. Panel (b) shows the initial GW frequency distribution of the selected CMC BBHs calculated from their component masses. The grey dashed line and number mark the median value.}\label{fig:data}
    \end{minipage}
\end{figure}

\begin{figure*}
    \centering
    \includegraphics[width=0.96\textwidth,
    trim=0 0 0 0,clip]{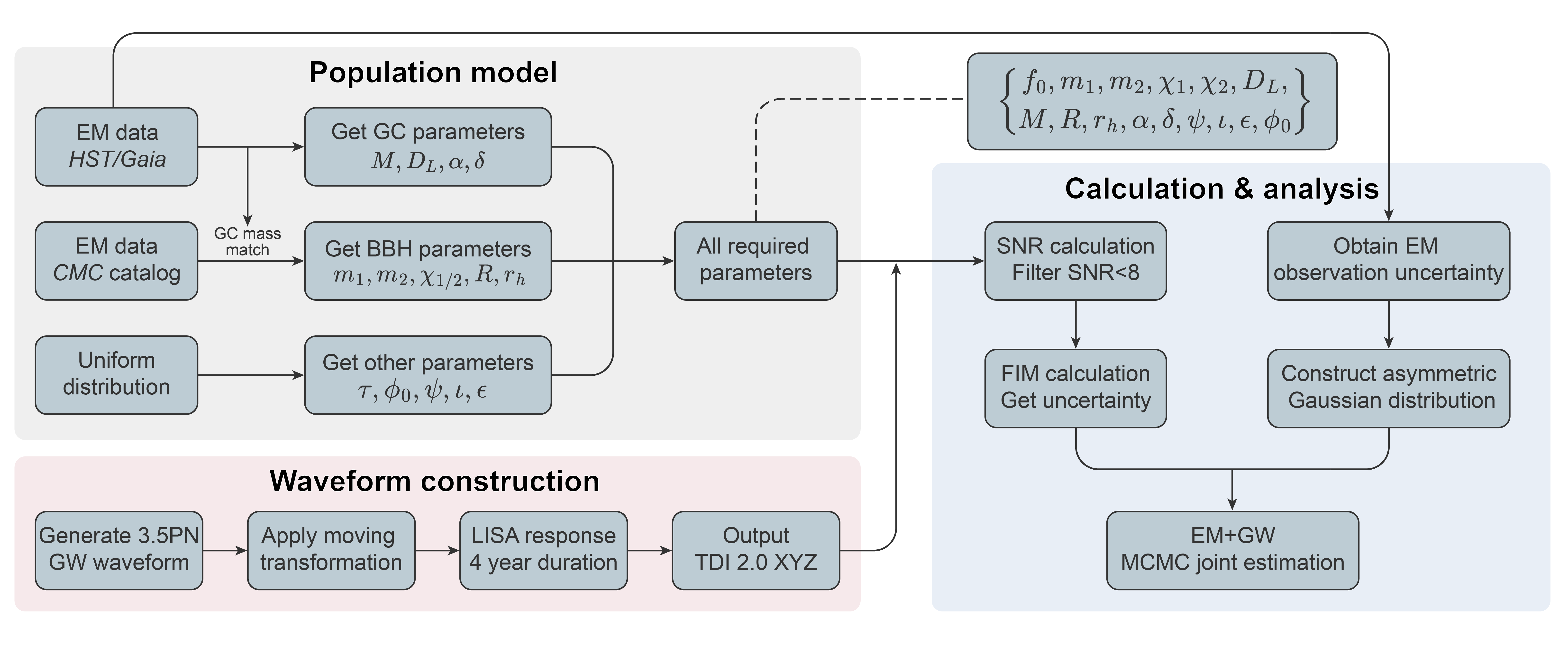}
    \caption{Schematic workflow of the data generation and analysis process. The pipeline consists of three main components: population modelling (top-left), waveform construction (bottom), and uncertainty estimation (right).}
    \label{fig:flowsheet}
\end{figure*}

For BBH properties, we utilize simulation data from \citet{CMC}, which provides a catalogue of large-scale cluster Monte Carlo (CMC) simulations.
These simulations, based on the Hénon-type Monte Carlo algorithm, incorporate a wide range of physical processes and follow the long-term dynamical evolution of dense star clusters. 
Figure~\ref{fig:data}(a) illustrates the number of BBHs contained in CMC clusters of different masses.
The total number of BBH mergers in a GC increases approximately linearly with the cluster mass, and the proportionality is affected by the initial virial radius $r_v$~\citep{Rodriguez2015}.  
This trend can be summarized by
\begin{equation}
    N_\mathrm{BBH} = k \left( \frac{M}{10^{5}\,\mathrm{M_\odot}} \right),
\end{equation}
where the coefficient $k$ depends on the choice of $r_v$~\citep{CMC}.  
These findings demonstrate that BBHs are a natural dynamical product of GC evolution, providing theoretical support for the use of BBH-generated GWs as observational probes of GC properties.

The CMC models allow BBHs to be dynamically ejected. We therefore use only the ``In-cluster two-body mergers'' as observational sources. These systems account for approximately 35\% of all CMC BBH mergers~\citep{CMC}.
Using Eq.~\eqref{eq:f0}, we calculate the initial GW frequencies of these BBHs, as shown in Fig.~\ref{fig:data}(b).
The inspiralling BBHs considered here evolve slowly in frequency, so the initial frequency provides a useful characterization of the source population.
Most of the selected BBHs have initial frequencies close to the most sensitive part of the LISA band, which is favourable for constraining GC parameters with GW observations.

For each EM-observed GC, we use the EM-inferred GC mass to select BBH entries from the CMC simulations.
The entries are ranked by the absolute mass difference $|M_{\rm obs}-M_{\rm CMC}|$ between the observed GC mass and the CMC host-cluster mass, and the 20 nearest entries are retained rather than imposing a fixed mass-difference threshold.
Here $M_{\rm CMC}$ is the CMC cluster total mass at the BBH merger time, obtained from the dynamical output of the corresponding CMC model rather than from the initial cluster mass.
The few-year to few-hundred-year offset between the LISA inspiral signal and merger is negligible compared with GC mass-evolution time-scales, so this merger-time mass provides an appropriate basis for assigning CMC BBH parameters to the EM-observed GCs.
The parameters extracted include the component masses ($m_1$ and $m_2$), the BBH spins ($\chi_1$ and $\chi_2$), the BBH outer orbital radius ($R$), and the GC half-mass radius ($r_h$) (see Fig.~\ref{fig:flowsheet}).

\begin{table}
\centering
\renewcommand{\arraystretch}{1.5}
\caption{Parameter distribution used in this study~\citep[see][]{LISA_SOBBH}. $U[a,b]$ represents a uniform distribution from $a$ to $b$.}\label{tab:parameters}
\begin{tabular*}{\columnwidth}{@{\extracolsep{\fill}}cc@{}}
\hline
 Parameter & Distribution \\
\hline
$\tau$  [yrs]
& $U[10,1000]/(1+z)$\\
$\phi_0,\psi$ [rad]
& $U[0,2\pi]$ \\
$\iota,\epsilon$ [rad]
& $\arccos(U[-1,1])$ \\
\hline
\end{tabular*}
\end{table}

The angular source parameters are sampled as summarized in Table~\ref{tab:parameters}.
For the inner BBH orbit, we set the eccentricity to zero when constructing the PN carrier waveform.
Dynamically assembled BBHs in GCs are expected to retain eccentricity in the LISA band~\citep{SamsingDOrazio2018LISA1,DOrazioSamsing2018LISA2}.
In this work, however, our focus is the GC-induced modulation from the BBH centre-of-mass motion rather than eccentric corrections to the inner-binary waveform.
Since the waveform impact of eccentricity is expected to become non-negligible mainly for sufficiently large eccentricities~\citep{Wu_PN}, we adopt the zero-inner-eccentricity approximation as a computationally convenient baseline.
The redshift $z$ is computed from the luminosity distance $D_L$ provided by EM observations, and the initial GW frequency $f_0$ is given by~\citep{Colpi_2017}
\begin{equation}
    \label{eq:f0}
    f_0 = \frac{1}{8\pi}
    \left(\frac{1}{5}\,\mathcal{M}_{c}^{5/3}\,\tau\right)^{-3/8},
\end{equation}
where $\mathcal{M}_{c} = (m_{1}m_{2})^{3/5}/(m_{1}+m_{2})^{1/5}$ is the chirp mass. 
We adopt the $\Lambda$CDM cosmological model with parameters provided by \citet{Planck_2018}, namely the Hubble constant $H_{0}=67.37~\mathrm{km\,s^{-1}\,Mpc^{-1}}$, the matter density parameter $\Omega_{m}=0.315$, and the dark energy density parameter $\Omega_{\Lambda}=0.685$.

This procedure yields a comprehensive set of source parameters for waveform modelling and uncertainty estimation.  
Considering the diversity of GC properties and the stochastic sampling of source parameters, the 20 nearest CMC BBH merger entries selected for each EM-observed GC provide a finite population sample for the FIM calculations while retaining the mass dependence of the CMC host clusters.  
For each of these BBHs, we compute the corresponding EM+GW constraint and statistically combine the results, thereby obtaining representative estimates of the achievable precision for each GC under realistic astrophysical conditions.

The parameter selection outlined above provides a complete specification of both the dynamical environment of the host GCs and the intrinsic properties of the BBH sources. These parameters form the basis for waveform construction and for assessing the impact of GC-induced modulation.
In the next section, we describe the data analysis procedure used to compute the SNR, construct the FIM, and quantify the resulting constraints on GC properties.

\subsection{Data analysis}\label{subsec:data_analysis}

The previous sections introduced the population modelling of BBHs in GCs and the construction of their modulated waveforms, establishing the physical inputs required for quantitative analysis.  
With these elements in place, this section describes the computational and statistical methods used to evaluate the detectability of each system and to infer the corresponding constraints on GC parameters.
The full analysis pipeline, summarized in Fig.~\ref{fig:flowsheet}, first evaluates the signal strength, then estimates parameter uncertainties using FIM, and finally incorporates EM information to obtain joint EM+GW constraints.

The analysis starts by computing the signal-to-noise ratio (SNR) for every BBH–GC system in the catalogue.
The SNR serves as the primary criterion for determining whether a signal is detectable by LISA and whether its modulation features can contribute to parameter inference.  
We exclude any source with $\mathrm{SNR} < 8$, adopting this as the detection threshold for the four-year observation~\citep{Tiwari2024,SNR_threshold}.
The SNR is given by~\citep{SNR}
\begin{equation}
\mathrm{SNR} = \sqrt{(h | h)},
\end{equation}
where the inner product $(a | b)$ is defined as
\begin{equation}
(a | b) = 4 \, \mathrm{Re} \int_0^{\infty} 
\frac{\tilde{a}^*(f)\tilde{b}(f)}{S_n(f)} df,
\end{equation}
where $S_n(f)$ is the one-sided PSD of the detector noise, and $\tilde{a}(f)$ and $\tilde{b}(f)$ are the Fourier transforms of $a(t)$ and $b(t)$, respectively.
Here, $S_n(f)$ denotes the one-sided noise PSD used in the TDI-channel inner product. The instrumental-noise PSD is constructed following \citet{Babak2021LISASensitivity} and \citet{Robson2019LISASensitivity}, while the four-year Galactic confusion foreground is treated following our previous work \citet{Wu_GB}. The source-dependent TDI response is retained separately in the waveform and FIM. Its relation to the PSD and the sensitivity curve is discussed in \citet{Wu_TDI,Wu_polar}.

For sources that satisfy the SNR threshold, parameter uncertainties are quantified through the FIM.  
This method offers an efficient estimate of the achievable measurement precision for LISA detection.  
The FIM is defined as~\citep{CutlerFlanagan1994,FIM}
\begin{equation}
\Gamma_{ij} = \left( \frac{\partial h}{\partial \theta_i} \middle| 
\frac{\partial h}{\partial \theta_j} \right),
\end{equation}
where $\theta_i$ denotes the $i$-th parameter in the waveform model. 
The inverse of $\Gamma$ yields the parameter covariance matrix, and the 1$\sigma$ uncertainty in $\theta_i$ is given by $\sigma_{\mathrm{GW}\, i} = \sqrt{(\Gamma^{-1})_{ii}}$.
This yields the GW constraints on the GC distance and mass for each detectable source, along with uncertainties in the extrinsic and intrinsic BBH parameters.
We check the numerical derivatives by repeating the FIM calculation with multiple finite-difference step sizes.
We then use the regime in which both the derivatives and the resulting parameter uncertainties reach a stable plateau.
The covariance matrix is computed following the matrix-inversion strategy adopted in~\citet{Iacovelli2022GWFAST}.
As an internal consistency check, we define
\begin{equation}
\epsilon_{\rm inv}=\max_{ij}
    \left|\left(\Gamma\Gamma^{-1}-I\right)_{ij}\right|,
\end{equation}
All FIM results retained in the analysis satisfy $\epsilon_{\rm inv}<10^{-4}$, which ensures that the numerical inversion is sufficiently stable for the uncertainty estimates used below.

While GW observations alone already constrain the GC mass and distance through modulation-induced features, the available EM measurements contain complementary information.  
To combine the two observations consistently, the GC parameters are modelled within a unified statistical framework.   
Since many EM-based uncertainties are significantly asymmetric, we model them using an asymmetric normal distribution $p_{\mathrm{EM}}(\theta)$, which provides a more faithful representation of the observational results~\citep{asym_error}.
The asymmetric normal distribution is expressed as
\begin{equation}
p_{\mathrm{EM}}(\theta) \propto 
\begin{cases}
\exp\!\left[-\dfrac{(\theta - \mu)^2}{2\sigma_1^2}\right], & \theta < \mu, \\[10pt]
\exp\!\left[-\dfrac{(\theta - \mu)^2}{2\sigma_2^2}\right], & \theta \ge \mu,
\end{cases}
\label{eq:asym_gauss}
\end{equation}
where $\mu$ denotes the central value, and $\sigma_1$ and $\sigma_2$ represent the lower and upper uncertainties, respectively~\citep{asym_error}.
Based on the uncertainties $\sigma_{\mathrm{GW}}$ derived from the FIM, we construct a corresponding normal distribution $p_{\mathrm{GW}}(\theta)$, adopting the same functional form as Eq.~\eqref{eq:asym_gauss}, with $\sigma_1 = \sigma_2 = \sigma_{\mathrm{GW}}$.

The joint EM+GW constraint is then obtained by multiplying the two independent distributions,
\begin{equation}
p_{\mathrm{EM+GW}}(\theta) \propto 
p_{\mathrm{EM}}(\theta)\, p_{\mathrm{GW}}(\theta).
\end{equation}
The GW contribution provides additional information that is independent of the EM dataset, so the width of the joint likelihood is generally narrower than that of $p_{\mathrm{EM}}(\theta)$ alone for the retained valid cases, reflecting an improvement in measurement precision.

We then numerically evaluate the joint distribution $p_{\mathrm{EM+GW}}(\theta)$ and obtain the resulting constraints on GC distance and mass from its central intervals.
Details of the asymmetric modelling procedure and numerical evaluation can be found in \citet{asym_error}.
Collectively, these steps provide a practical and theoretically grounded framework for extracting improved GC parameters via multi-messenger observations.

\subsection{Plummer Enclosed-Mass Relation and Fisher Priors}\label{subsec:priori}

The BBH outer orbit is governed by the mass enclosed within its orbital radius, not necessarily by the total GC mass.
To connect these two quantities, we model the smooth GC potential with a Plummer profile, a simple spherical model with a finite central density and a rapidly declining outer density~\citep{Plummer1911}.
The corresponding density profile is~\citep{BinneyTremaine2008}
\begin{equation}
\rho_{\rm GC}(r)=\frac{3M}{4\pi b^3}\left(1+\frac{r^2}{b^2}\right)^{-5/2},
\end{equation}
where $M$ is the total GC mass and $b$ is the Plummer scale radius.
The three-dimensional half-mass radius satisfies $r_h \simeq 1.305 b$.
For a BBH with outer orbital radius $R$, the enclosed mass is then
\begin{equation}
M_{\rm enc}(R)=M\frac{R^3}{(R^2+b^2)^{3/2}}.
\label{eq:menc}
\end{equation}
This relation is used in the orbital velocity and modulation calculations, while $M$ remains the total GC mass constrained by EM and GW observations.
The CMC simulations adopt King profiles for the initial cluster models~\citep{CMC}. For analytical simplicity and efficient Fisher-matrix calculations, we approximate the smooth GC potential with the Plummer profile~\citep{Plummer1911}. After matching the same total mass $M$ and three-dimensional half-mass radius $r_h$, the two profiles differ only modestly over the BBH outer-orbit radii relevant to our analysis; a quantitative comparison is given in Appendix~\ref{app:plummerking}.

\begin{figure}
    \begin{minipage}{\columnwidth}
        \centering
        \includegraphics[width=0.96\textwidth,
        trim=0 0 0 0,clip]{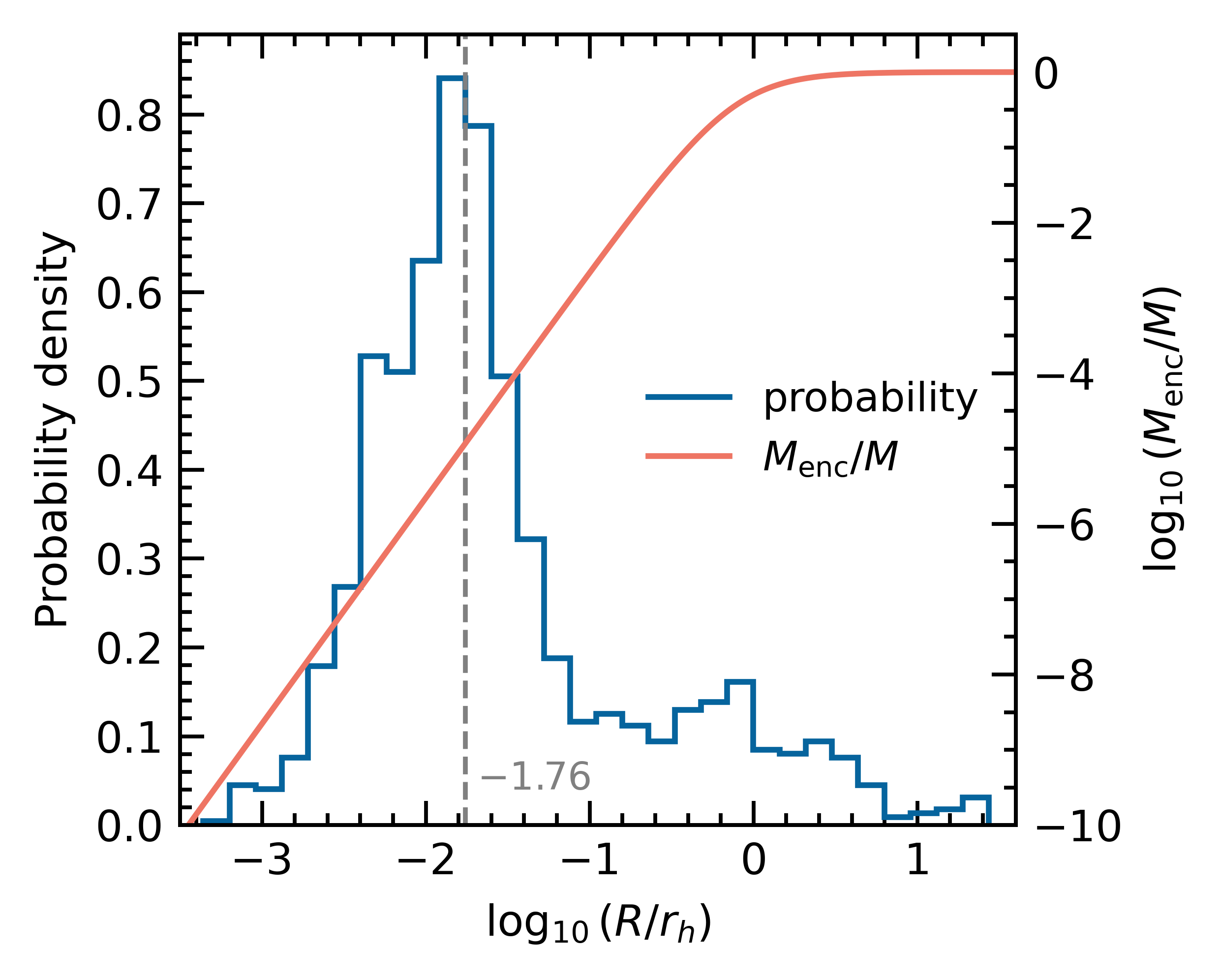}
        \caption{Distribution of the ratio between the BBH outer orbital radius $R$ and the GC half-mass radius $r_h$ for the selected CMC BBHs. The histogram shows the distribution of $R/r_h$, while the solid curve shows the enclosed-mass fraction $M_{\rm enc}/M$ computed from the Plummer profile. The dashed line and number mark the median value.}\label{fig:radius}
    \end{minipage}
\end{figure}

Figure~\ref{fig:radius} shows that most BBH outer orbits lie well inside the GC half-mass radius.
The median of the selected CMC population is $R\simeq 1.74\times10^{-2}r_h$, corresponding to $M_{\rm enc}\simeq 1.17\times10^{-5}M$.
The enclosed mass can therefore be much smaller than the total GC mass, reaching fractions from about $10^{-2}$ down to $10^{-7}$ for the relevant part of the CMC population.
Using the total GC mass directly as the enclosed orbital mass would overestimate the orbital velocity and lead to overly optimistic parameter constraints.

The Plummer relation also introduces a correlation between $R$ and $r_h$ in the FIM calculation.
We therefore include a weak population-informed Gaussian prior on their ratio, $\eta=R/r_h$.
The prior is calibrated from the selected CMC BBHs that merge inside the cluster and have finite outer-orbit radii.
Specifically, we use the central 68\% interval of the CMC distribution to define
\begin{equation}
\ln\eta_0=\ln\eta_{50},\qquad
\sigma_{\ln\eta}=\frac{1}{2}\left(\ln\eta_{84}-\ln\eta_{16}\right),
\end{equation}
where $\eta_{16}$ , $\eta_{50}$ , and $\eta_{84}$ are the 16th, 50th, and 84th percentiles, respectively.
Following the standard Fisher treatment of Gaussian priors as additive prior matrices~\citep{Iacovelli2022GWFAST}, the total FIM is
\begin{equation}
\Gamma_{\rm total}=\Gamma_{\rm GW}+\Gamma_{\rm prior}.
\end{equation}
The prior acts on a parameter combination, so we project it onto the Fisher parameters through the corresponding variable transformation~\citep{Albrecht2009JDEM}.
Since both $R$ and $r_h$ are treated as logarithmic FIM parameters,
\begin{equation}
\frac{\partial\ln\eta}{\partial\ln R}=1,\qquad
\frac{\partial\ln\eta}{\partial\ln r_h}=-1.
\end{equation}
Thus the prior contributes the following sub-block in the $(\ln r_h,\ln R)$ sector:
\begin{equation}
\Gamma_{\rm prior}
=\frac{1}{\sigma_{\ln\eta}^{2}}
\begin{pmatrix}
1 & -1\\
-1 & 1
\end{pmatrix}.
\end{equation}
This prior constrains only the relative scale $R/r_h$. It does not fix either $R$ or $r_h$ individually.
With this enclosed-mass prescription and ratio prior, the waveform calculation consistently links the total GC mass constrained by EM and GW data to the local mass that controls the BBH outer orbit.
$M$, $R$, and $r_h$ then define a complete dynamical description of the GC potential and the BBH outer orbit for the uncertainty estimates presented in the next section.
\FloatBarrier

\section{Results}\label{sec:Results}

\begin{figure*}
    \centering
    \includegraphics[width=0.96\textwidth,
    trim=0 0 0 0,clip]{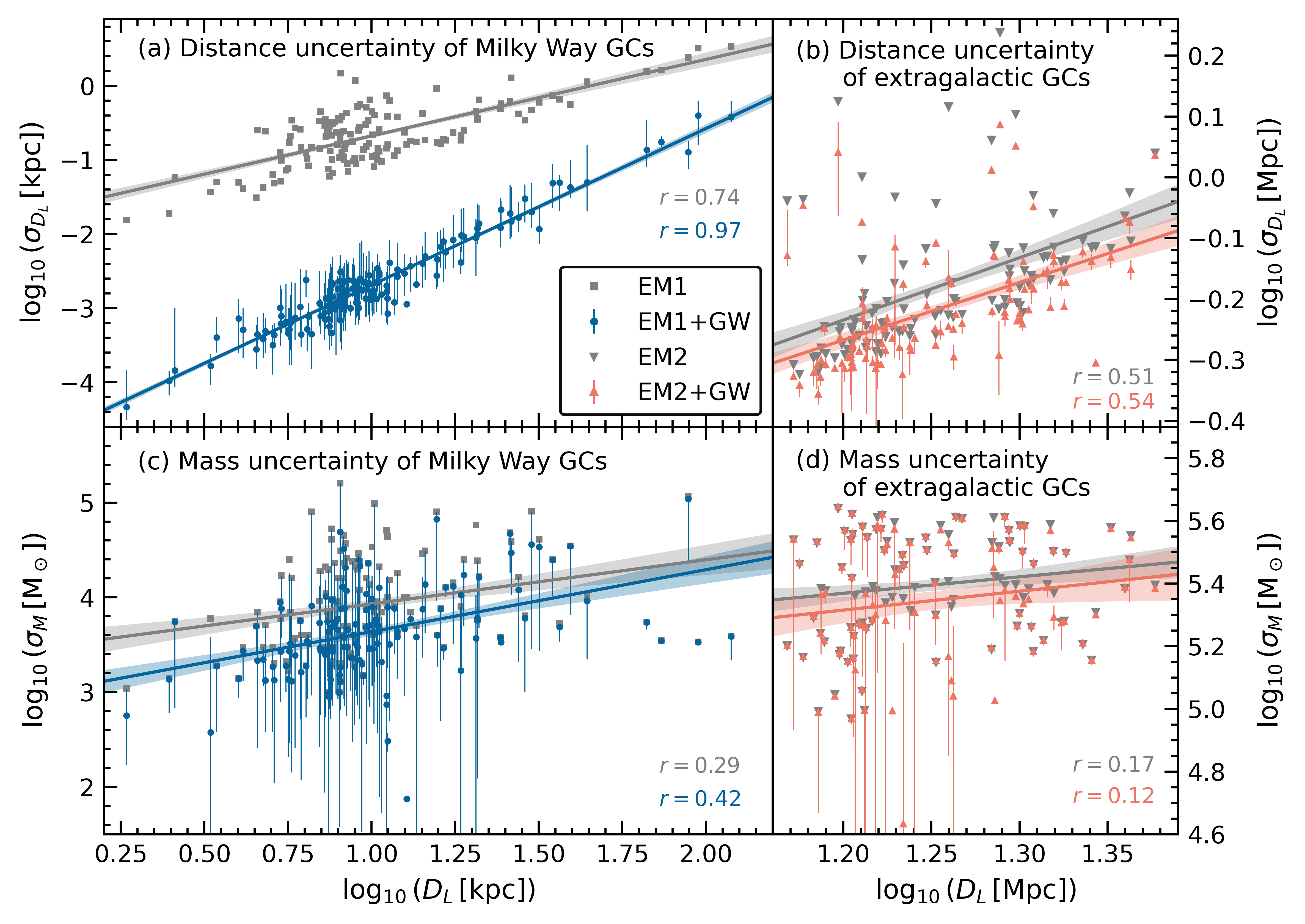}
    \caption{Relationship between the uncertainty of GC parameters and distance. Panels (a) and (b) show the distance uncertainties for Milky Way and extragalactic GCs, respectively, while panels (c) and (d) show the mass uncertainties. Grey markers represent existing EM-only measurements, and coloured markers denote the results after incorporating GW observations (EM+GW). Vertical bars indicate the $1\sigma$ spread in uncertainties arising from different BBH configurations within each GC. The reported value $r$ refers to the Pearson correlation coefficient. Both axes use a logarithmic scale. Solid lines show linear fits to the data in logarithmic space.}
    \label{fig:distance_error}
\end{figure*}

To quantify the improvement introduced by modulated GW signals, we evaluate the uncertainties of the relevant parameters using the FIM.
We begin with two benchmark samples constructed from existing EM observations: one for Milky Way GCs and one for extragalactic GCs. The corresponding EM data sources are summarized in Appendix~\ref{app:emdata}.
For each EM-observed GC, we assign the 20 nearest CMC BBH entries according to the mass-matching procedure described in Sec.~\ref{subsec:param_selection}, and combine the valid GW uncertainty estimates with the EM uncertainties following Sec.~\ref{subsec:data_analysis}.
All results discussed below are restricted to cases that pass the SNR threshold and the FIM inversion-stability criterion. Consequently, some EM-observed GCs, or some of the 20 CMC BBH matches assigned to a given GC, may be excluded from the plotted samples.
Therefore, the plotted GW-assisted samples should be interpreted as the subset of EM-observed GCs and CMC matches for which the simulated GW signal is detectable and yields a stable FIM estimate.

Figure~\ref{fig:distance_error} shows the reduction in GC parameter uncertainties after adding GW information to existing EM measurements.
Panels (a) and (b) show that the distance uncertainty generally increases with $D_L$, with the clearest positive correlation appearing in the Milky Way GC sample.
This trend follows from the inverse dependence of GW amplitude on distance. More distant sources have weaker GW signals and therefore provide less information in the FIM~\citep{Wu_GB}.
This behaviour is also reflected by the fitted correlations, with the EM+GW distance uncertainty showing a strong correlation for Milky Way GCs and a weaker but still positive correlation for extragalactic GCs.
For Milky Way GCs, the median relative distance uncertainty decreases from $2.64\%$ for EM-only measurements to $0.03\%$ after including GW information.
For extragalactic GCs, the improvement is much more limited, with the median relative uncertainty decreasing only from $3.78\%$ to $3.45\%$.

Panels (c) and (d) show the corresponding mass uncertainties. Compared with the distance uncertainty, the mass uncertainty has a weaker correlation with $D_L$, indicating that mass information is not controlled only by the overall GW amplitude or signal strength.
In our model, the GC mass affects the waveform through the frequency modulation induced by the BBH outer orbit. The mass constraint therefore also depends on the outer-orbit radius, the GC half-mass radius, and the enclosed mass that sets the dynamical modulation strength~\citep{Wu_SMBH}.
For the Milky Way sample, the median relative mass uncertainty decreases from $10.19\%$ to $6.13\%$, showing a moderate improvement for nearby systems.
For the extragalactic sample, the median relative mass uncertainty decreases from $48.52\%$ to $45.43\%$, so the overall improvement remains limited.
Thus, the gain in mass measurement should be interpreted as most effective for nearby or strongly modulated BBH configurations, rather than as a uniform improvement across all GCs.

\begin{figure}
    \begin{minipage}{\columnwidth}
        \centering
        \includegraphics[width=0.96\textwidth,
        trim=0 0 0 0,clip]{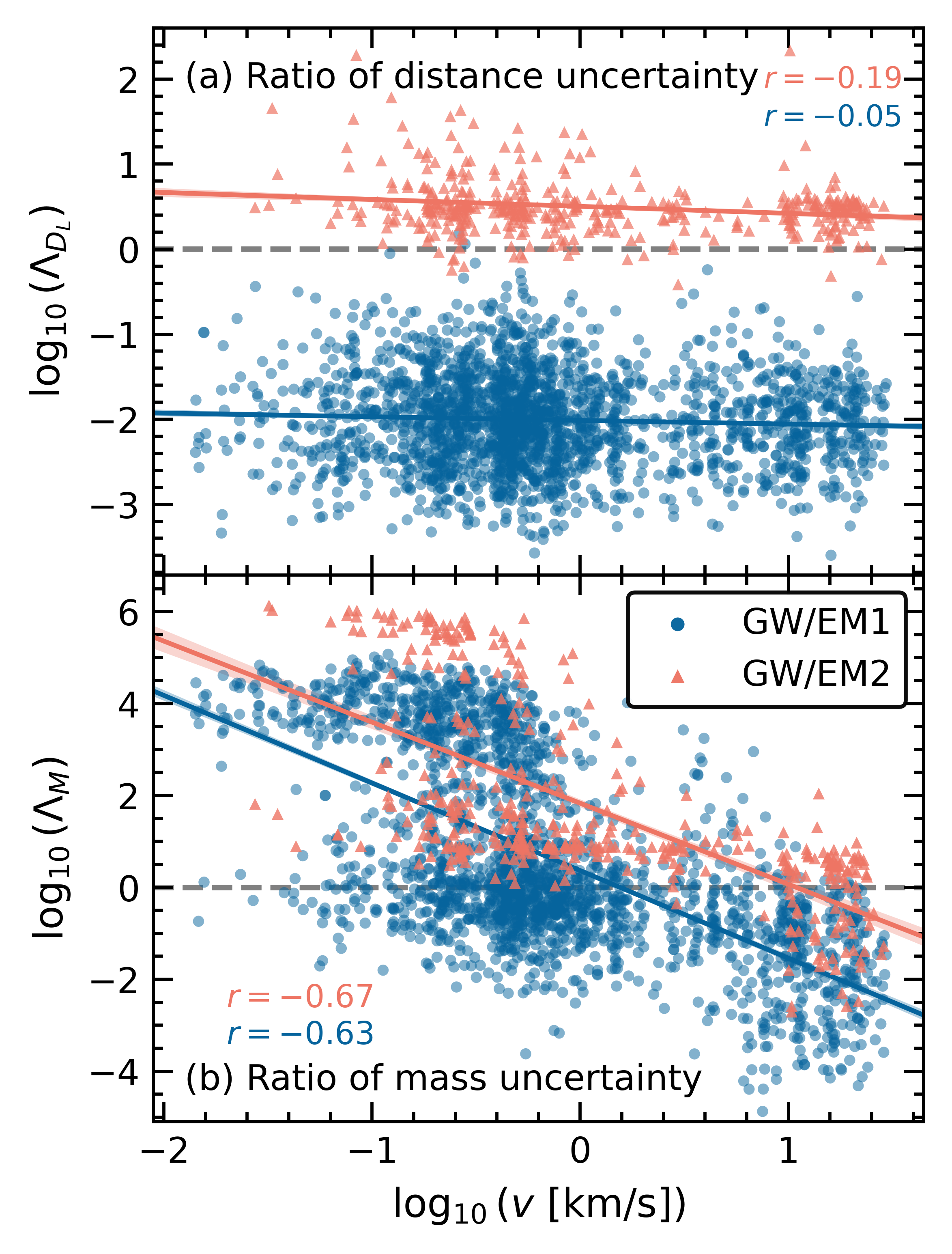}
        \caption{Relationship between the uncertainty ratio of GC parameters and the BBH outer-orbit velocity. Panels (a) and (b) show the ratios for distance and mass, respectively, defined as $\Lambda=\sigma_{\rm GW}/\sigma_{\rm EM}$. Each marker corresponds to one valid EM-observed GC--CMC BBH match among the 20 nearest CMC matches assigned in Sec.~\ref{subsec:param_selection}. The grey dashed line marks $\Lambda=1$, where the GW and EM uncertainties are equal, $\sigma_{\rm GW}=\sigma_{\rm EM}$. Other plot elements follow the same style as in Fig.~\ref{fig:distance_error}.}\label{fig:velocity_error}
    \end{minipage}
\end{figure}

In addition to the overall amplitude dependence on distance, the strength of waveform modulation is closely linked to the orbital velocity of the BBH around the GC.
To separate distance-related amplitude effects from dynamical modulation by the outer orbit, we examine the uncertainty ratio $\Lambda=\sigma_{\rm GW}/\sigma_{\rm EM}$ as a function of the BBH outer-orbit velocity in Fig.~\ref{fig:velocity_error}.
Each point in this figure corresponds to one valid EM-observed GC--CMC BBH match, rather than to the median of the 20 matches assigned to an individual GC.
Values of $\Lambda<1$ indicate that the GW uncertainty is smaller than the EM uncertainty, whereas $\Lambda>1$ indicates that the GW constraint is weaker than the EM constraint.

Figure~\ref{fig:velocity_error}(a) shows that the distance ratio $\Lambda_{D_L}$ has little clear correlation with the BBH outer-orbit velocity.
This behaviour indicates that the distance measurement is mainly controlled by the GW amplitude and the overall signal strength, not directly by the outer-orbit velocity.
Most Milky Way GC matches lie below $\Lambda_{D_L}=1$, showing that GW observations can often improve the distance constraints for nearby sources.
Many extragalactic GC matches instead remain above $\Lambda_{D_L}=1$, consistent with their weaker signals at larger distances.
Panel (b) shows a different behaviour for the mass ratio $\Lambda_M$, which decreases clearly as the BBH outer-orbit velocity increases.
At low velocities, many matches have $\Lambda_M>1$, so the GW mass constraint is weaker than the EM constraint. At higher velocities, more matches enter the $\Lambda_M<1$ region, indicating that stronger outer-orbit modulation can make the GW mass constraint more informative.
This contrast shows that distance measurements are dominated by amplitude information, while mass measurements are more sensitive to the dynamical modulation imprinted by the BBH outer motion~\citep{Wu_kick,Wu_SMBH}.
The fitted correlations support this interpretation: the distance ratios are nearly uncorrelated with velocity, whereas the mass ratios show a clear negative correlation with velocity for both GC samples.

The velocity dependence can be understood from the outer-orbit dynamics. In a Plummer potential, the BBH outer-orbit velocity is controlled by the orbital radius $R$ and the enclosed mass $M_{\rm enc}(R)$, with the scaling relation
\begin{equation}
v\simeq\sqrt{M_{\rm enc}(R)/R}.
\label{eq:outer_orbit_velocity}
\end{equation}
Here $M_{\rm enc}(R)$ is determined by the total GC mass $M$, the half-mass radius $r_h$, and the BBH outer-orbit radius $R$.
A larger enclosed mass or a more compact outer orbit produces a larger velocity, which strengthens the Doppler and relativistic frequency modulation of the GW signal.
Since GWs are highly sensitive to frequency evolution, stronger velocity-induced modulation makes the mass-related information easier to extract.
This explains why Fig.~\ref{fig:velocity_error} shows a clear velocity dependence for the mass ratio but only a weak velocity dependence for the distance ratio.

Overall, these results show that the contribution of GW modulation to GC parameter estimation is conditional.
For nearby Milky Way GCs, the stronger signals allow GW observations to improve the distance measurement substantially and the mass measurement moderately.
For extragalactic GCs, the larger distances limit the information carried by the GW signal, so the overall improvement in both distance and mass is modest.
The mass improvement also depends strongly on the BBH outer-orbit velocity, and GW constraints become most competitive when the outer-orbit modulation is sufficiently strong.
The method is therefore best suited to nearby GCs and BBH configurations with strong outer-orbit modulation, rather than providing the same level of improvement for all GC samples.
\section{Discussion}\label{sec:Discussion}

In this work, we assume that the host GC remains stationary and that the BBH follows a stable outer orbit throughout the observation period.
For GW detection, the observation duration is set to four years.
Given that the typical proper motion of GCs is on the order of mas/yr, the resulting positional shifts are negligible and do not influence the GW response (as it is computed based on the source's sky location)~\citep{proper_motions}.
Although the GC’s radial velocity can be comparable in magnitude to the BBH’s orbital velocity, it remains effectively constant in both direction and speed~\citep{radial_velocity1,radial_velocity2}.
As a result, it induces only a static phase offset in the GW waveform—degenerate with the initial phase—and thus does not lead to any modulation~\citep{Cao_GWtrans,Wu_kick}.
In a real detection, the association between a modulated BBH signal and a GC would require additional validation rather than being inferred from the velocity alone.
A practical strategy would combine the GW-inferred sky localization and modulation pattern with EM information on candidate GCs, including their positions, distances, masses, and structural scales.
The BBH--GC association would then be supported if the inferred orbital velocity and enclosed mass are consistent with the potential of the candidate GC.
Such host-association tests are beyond the scope of the present FIM forecast, where the BBH--GC association is assumed when constructing the simulated source population.

On longer time-scales, several astrophysical processes—such as Lense-Thirring precession, Lidov-Kozai oscillations, and dynamical friction from ambient gas—could potentially alter the outer orbit~\citep{orbit_effect1,orbit_effect2,orbit_effect3}.
However, \citet{orbit_effect4} show that these effects typically operate on time-scales ranging from thousands to millions of years, far exceeding the duration of observation, and also assume a circular inner orbit for the stellar-mass BBH.
Therefore, the assumption of a stable outer orbit is well justified and enables a tractable yet physically meaningful model for our analysis.
Following this circular-inner-orbit baseline, we set the inner eccentricity to zero. Dynamically formed GC BBHs may retain LISA-band eccentricity~\citep{SamsingDOrazio2018LISA1,DOrazioSamsing2018LISA2}, but eccentricity mainly modifies the carrier waveform and is expected to have a non-negligible effect here only when it is sufficiently large~\citep{Wu_PN}.

Building on these assumptions, we employed a FIM analysis to estimate the parameter uncertainties of BBHs in GCs using GW observations.
Combined with existing EM measurements, we constructed asymmetric normal likelihoods to model the joint EM+GW constraints in a multi-messenger framework.
This establishes a complete methodology for using BBH-generated GWs as dynamical probes to infer GC distance and mass.
Our results demonstrate that this approach can substantially improve distance constraints for nearby Milky Way GCs and can improve mass constraints when the BBH outer-orbit modulation is sufficiently strong.
The Plummer profile used here provides a simple spherical mapping between the total GC mass and the enclosed mass at the BBH outer orbit. Real GCs may have more complex density profiles, so future applications should test how the inferred constraints depend on the adopted cluster potential model.

It is important to note that the GW-based constraints presented in this study are derived from theoretical estimates using the FIM.
Our aim is to theoretically validate the feasibility of this method.
Given the high computational cost of Bayesian inference in a 15-dimensional parameter space, we employ the FIM to estimate parameter uncertainties.
In the future, we plan to carry out such analyses using real space-based GW data. 
This will allow us to further test the practical viability of this technique and eventually perform full Bayesian inference to extract GC properties from GW signals.

\section*{Acknowledgements}

This work was supported by the National Key Research and Development Program of China (grant number 2023YFC2206702), the National Natural Science Foundation of China (grant numbers 125B2102, 12575072, and 12547101), the Fundamental Research Funds for the Central Universities Project (grant number 2024IAIS-ZD009), and the Natural Science Foundation of Chongqing (grant number CSTB2023NSCQ-MSX0103).

\section*{Data Availability}

The EM observational data of Milky Way GCs are taken from~\citet{MWGC}.
The extragalactic GC sample is constructed from the ACSVCS and ACSFCS catalogues, together with host galaxy SBF distances~\citep{Jordan2009ACSVCS,Jordan2015ACSFCS,Blakeslee2009SBF}.
The corresponding catalogues are publicly available at
\url{https://people.smp.uq.edu.au/HolgerBaumgardt/globular/},
\url{https://vizier.cds.unistra.fr/viz-bin/VizieR?-source=J/ApJS/180/54}, and
\url{https://vizier.cds.unistra.fr/viz-bin/VizieR?-source=J/ApJS/221/13}.
The CMC simulation data employed in this study are obtained from~\citet{CMC} and can be accessed at  
\url{https://cmc.ciera.northwestern.edu/home/}.



\bibliographystyle{mnras}
\bibliography{example} 


\appendix

\section{Electromagnetic data}\label{app:emdata}

This appendix summarizes the EM GC data used in this work.
For Milky Way GCs, we use the same input catalogue as in the original analysis, which provides the cluster mass, luminosity distance, sky position, and corresponding observational uncertainties~\citep{MWGC}.
For extragalactic GCs, we use individual GC photometry from the ACS Virgo Cluster Survey and the ACS Fornax Cluster Survey~\citep{Jordan2009ACSVCS,Jordan2015ACSFCS}.

The luminosity distance of each extragalactic GC is assigned from the surface brightness fluctuation distance of its host galaxy.
We use the host galaxy distance moduli tabulated for the Virgo and Fornax samples by~\citet{Blakeslee2009SBF} and compute
\begin{equation}
D_L({\rm Mpc})=10^{(\mu-25)/5}.
\end{equation}
The distance uncertainty is propagated from the uncertainty in the distance modulus as
\begin{equation}
\sigma_{D_L}=\frac{\ln 10}{5}D_L\sigma_{\mu}.
\end{equation}
Thus, the extragalactic distances used here are host specific SBF distances rather than typical Virgo or Fornax mean distances.

The ACSVCS and ACSFCS catalogues do not directly provide individual GC masses.
We therefore estimate the mass of each extragalactic GC from its extinction-corrected F850LP luminosity.
The corrected magnitude is
\begin{equation}
z_0=z_{\rm mag}-1.485\,E(B-V),
\end{equation}
where the ACS/F850LP extinction correction follows the ACS photometric calibration used for these surveys~\citep{Sirianni2005ACS,Blakeslee2009SBF}.
The absolute magnitude and luminosity are then
\begin{equation}
M_z=z_0-\mu,\qquad
\log_{10}(L_z/L_{z,\odot})=-0.4(M_z-M_{z,\odot}).
\end{equation}
We adopt $M_{z,\odot}=4.51$ for the solar absolute magnitude in the AB $z$ band~\citep{Willmer2018Solar}, and use a fiducial constant mass-to-light ratio $\Upsilon_z=(M/L)_z=1.5$ with a systematic scatter of $\sigma_{\log_{10}\Upsilon}=0.20$ dex, consistent with the level expected for GC mass-to-light ratios~\citep{Baumgardt2017}.
The estimated mass is
\begin{equation}
M_{\rm GC}=\Upsilon_z L_z.
\end{equation}

The mass uncertainty includes the photometric uncertainty, the distance modulus uncertainty, and the adopted mass-to-light ratio scatter.
With $\sigma_{E(B-V)}=0.01~{\rm mag}$, the magnitude uncertainty entering the luminosity estimate is
\begin{equation}
\sigma_{z_0}^2=\sigma_z^2+\left(1.485\,\sigma_{E(B-V)}\right)^2,
\end{equation}
and the logarithmic mass uncertainty is
\begin{equation}
\sigma_{\log_{10}M}^2=
\sigma_{\log_{10}\Upsilon}^2
+(0.4\sigma_{z_0})^2
+(0.4\sigma_\mu)^2.
\end{equation}
The corresponding linearized mass uncertainty is obtained through $\sigma_M=\ln(10)M_{\rm GC}\sigma_{\log_{10}M}$.

The final extragalactic sample is selected from candidates with $p_{\rm GC}\ge 0.90$, available $z_{\rm mag}$ and $E(B-V)$, and a matched host-galaxy SBF distance.
We additionally require the inferred mass to lie within the merger time CMC host cluster mass range available in the local CMC data used here.
The final EM sample contains 161 Milky Way GCs and 122 extragalactic GCs, including 83 Virgo GCs from ACSVCS and 39 Fornax GCs from ACSFCS.

\section{Comparison of Plummer and King profiles}\label{app:plummerking}

The CMC simulations adopt King profiles with \(W_0=5\) as their initial cluster models~\citep{CMC}. We therefore use \(W_0=5\) as the primary King model for assessing the Plummer approximation adopted in this work. To examine whether the comparison depends sensitively on the King concentration, we additionally consider two lower-concentration models with \(W_0=3\) and \(4\)~\citep{King1966}.
In the King model, the dimensionless central potential is defined by
\begin{equation}
\displaystyle
W_0\equiv\frac{\Psi(0)}{s_{\rm K}^2},
\end{equation}
where \(\Psi(0)\) is the relative gravitational potential at the GC centre and \(s_{\rm K}\) is the King velocity scale. The King model is a lowered-isothermal model with a finite tidal boundary. Its distribution function is~\citep{King1966,BinneyTremaine2008}
\begin{equation}
\displaystyle
f(\mathcal{E})=
\begin{cases}
A\left[\exp\!\left(\mathcal{E}/s_{\rm K}^{2}\right)-1\right], & \mathcal{E}>0,\\
0, & \mathcal{E}\leq0,
\end{cases}
\end{equation}
where \(A\) is a normalization constant. The relative energy and relative potential are defined by
\begin{equation}
\displaystyle
\mathcal{E}=\Psi(r)-\frac{v^2}{2}, \qquad \Psi(r)=\Phi(r_t)-\Phi(r).
\end{equation}
Here, \(v\) is the stellar speed, \(\Phi\) is the gravitational potential, and \(r_t\) is the tidal radius. The density profile is
\begin{equation}
\displaystyle
\rho^{\rm K}(r)=\int f(\mathcal{E})\,\mathrm{d}^{3}v.
\end{equation}
The density profile and relative potential are then determined self-consistently by combining the distribution function with the Poisson equation
\begin{equation}
\displaystyle
\nabla^{2}\Psi=-4\pi\rho^{\rm K}.
\end{equation}
For each \(W_0\), the King and Plummer models are normalized to the same total cluster mass \(M\) and three-dimensional half-mass radius \(r_h\). The King enclosed mass used in the comparison is therefore
\begin{equation}
\displaystyle
M_{\rm enc}^{\rm K}(R)=4\pi\int_0^R\rho^{\rm K}(r)r^2\,\mathrm{d}r,
\end{equation}
where \(R\) is the BBH outer-orbit radius. The corresponding Plummer enclosed mass \(M_{\rm enc}^{\rm P}(R)\) is given by Eq.~\eqref{eq:menc}.

\begin{figure}
    \begin{minipage}{\columnwidth}
        \centering
        \includegraphics[width=0.92\textwidth,
        trim=0 0 0 0,clip]{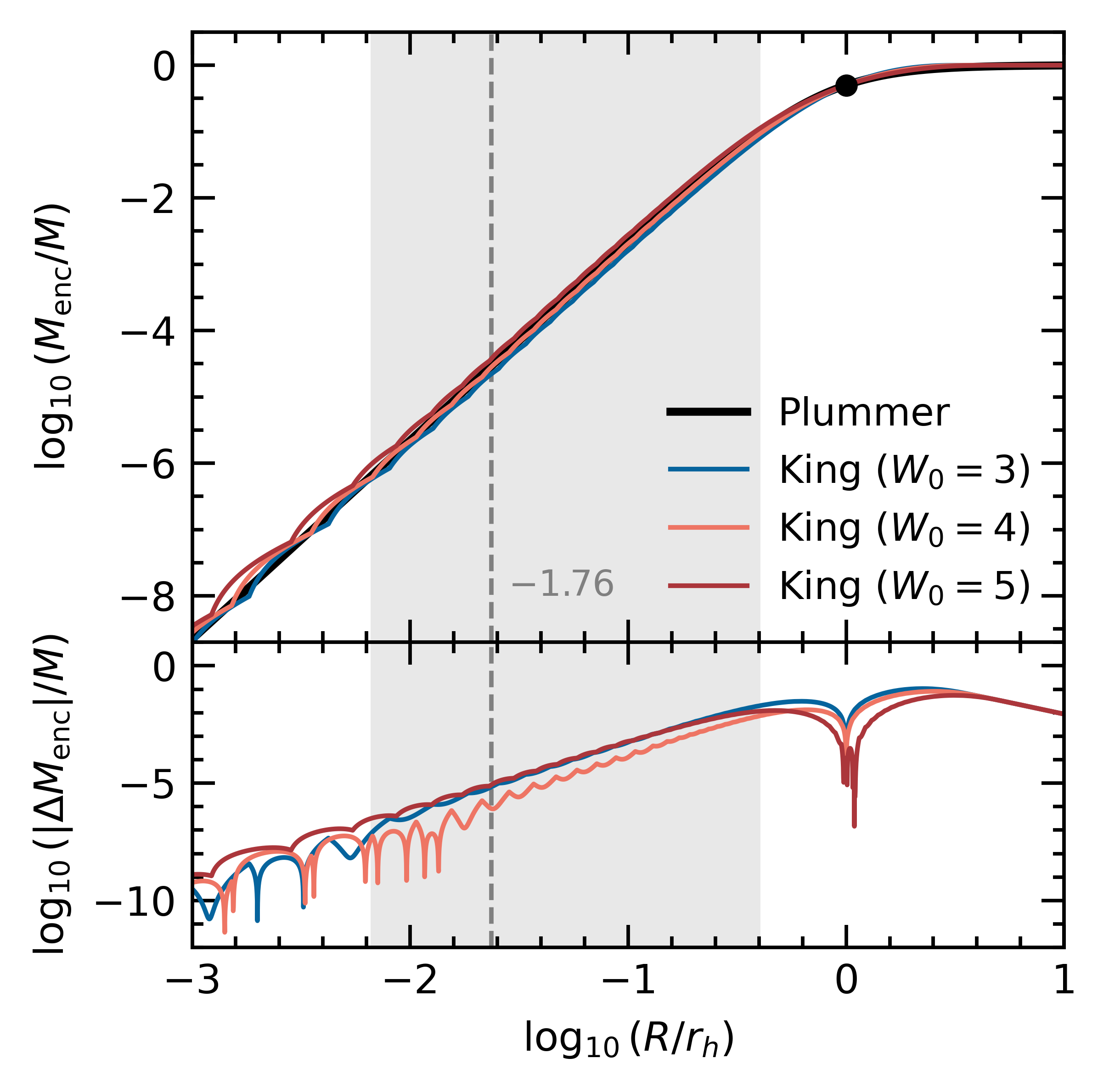}
        \caption{Comparison of the enclosed-mass profiles after matching the total cluster mass and three-dimensional half-mass radius. The upper panel shows the Plummer result \(M_{\rm enc}^{\rm P}/M\) and the King results \(M_{\rm enc}^{\rm K}/M\) for three values of \(W_0\). The shaded vertical band marks the 16th--84th percentile range of \(R/r_h\) for the valid CMC BBH sample, and the dashed vertical line marks its median. The lower panel shows the difference \(\Delta M_{\rm enc}/M\), where \(\Delta M_{\rm enc}=M_{\rm enc}^{\rm P}-M_{\rm enc}^{\rm K}\).}
        \label{fig:plummerking}
    \end{minipage}
\end{figure}

Figure~\ref{fig:plummerking} compares the enclosed-mass profiles of the Plummer model and King models with different \(W_0\). The profiles differ only modestly, particularly over the shaded range containing most of the valid CMC BBH sample. This agreement supports the use of the Plummer approximation for the outer orbits considered in this work.

To further compare the Plummer and King models, we consider their effects on the BBH outer-orbit velocity. The enclosed mass determines this velocity through Eq.~\eqref{eq:outer_orbit_velocity}. Therefore, the ratio of the BBH outer-orbit velocities obtained from the Plummer and King profiles provides a direct estimate of the possible change in the Doppler modulation induced by the choice of cluster potential. We quantify this ratio with
\begin{equation}
\displaystyle
\eta_v=\frac{v^{\rm P}}{v^{\rm K}}=
\left[\frac{M_{\rm enc}^{\rm P}(R)}{M_{\rm enc}^{\rm K}(R)}\right]^{1/2},
\end{equation}
where \(v^{\rm P}\) and \(v^{\rm K}\) are the Plummer and King BBH outer-orbit velocities, respectively. For parameters constrained primarily through the velocity-dependent moving-source modulation, and assuming that the corresponding uncertainty scales approximately inversely with the orbital-velocity amplitude, the profile-induced change can be characterized at the order-of-magnitude level by \(\sigma^{\rm K}\sim\eta_v\sigma^{\rm P}\), where \(\sigma^{\rm P}\) and \(\sigma^{\rm K}\) denote the uncertainties obtained with the Plummer and King profiles, respectively. This scaling should only be interpreted as an order-of-magnitude estimate for the possible impact on parameter uncertainties and does not replace a full Fisher-matrix calculation.

\begin{table}
\centering
\renewcommand{\arraystretch}{1.5}
\caption{The 16th percentile, median, and 84th percentile of the velocity ratio \(\eta_v=v^{\rm P}/v^{\rm K}\) for the valid BBH sample.}
\label{tab:velocityratio}
\begin{tabular*}{\columnwidth}{@{\extracolsep{\fill}}cccc@{}}
\hline
\(W_0\) & 16th & median & 84th\\
\hline
\(3\) & \(1.011\) & \(1.130\) & \(1.157\)\\
\(4\) & \(0.966\) & \(1.016\) & \(1.033\)\\
\(5\) & \(0.843\) & \(0.873\) & \(0.947\)\\
\hline
\end{tabular*}
\end{table}

Table~\ref{tab:velocityratio} gives the distribution of \(\eta_v\) over the valid BBH sample and quantifies the relative change in the orbital-velocity scale. Values of \(\eta_v\) close to unity indicate that the Plummer approximation introduces only a small dynamical difference compared with the corresponding King model. After matching the total cluster mass and three-dimensional half-mass radius, the Plummer approximation reproduces the enclosed-mass distribution of King models reasonably well within the BBH orbital region. The resulting variation in the orbital-velocity scale suggests that the effect of the adopted Plummer approximation on the waveform modulation is limited.



\bsp	
\label{lastpage}
\end{document}